\documentclass{raa_twocolumn}
\usepackage{graphicx,times}             
\usepackage{natbib}
\usepackage{amssymb,amsmath}
\bibpunct{(}{)}{;}{a}{}{,}
\usepackage{hyperref}
\usepackage{enumitem}

\begin{document}

  \title{In-flight performance of the MXT Camera
}

   \volnopage{Vol.0 (202x) No.0, 000--000}      
   \setcounter{page}{1}          

   \author{M.~Moita 
      \inst{1}
   \and A.~Meuris
      \inst{1}
   \and P.~Ferrando
      \inst{1}
    \and A.~Sauvageon
      \inst{1}
    \and H.~Goto
      \inst{1,2}
    \and D.~Götz
      \inst{1}
    \and C.~Plasse
      \inst{1}
    \and L.~Godinaud
      \inst{1}
    \and A.~Fort
\inst{3}
    \and K.~Mercier
\inst{3}
    \and S.~Crepaldi
\inst{3}
   }


   \institute{
             AIM-CEA/DRF/Irfu/Departement d’Astrophysique, CNRS, Universite Paris-Saclay, Universite Paris Cite, Orme des Merisiers, Bat. 709, Gif-sur-Yvette, 91191, France;\\
        \and
             College of Science and Engineering, School of Mathematics and Physics, Kanazawa University, Kakuma, 9201192 Kanazawa, Japan;\\
             \and
             Centre National d’Etudes Spatiales, Centre spatial de Toulouse, 18 avenue Edouard Belin, 31401 Toulouse Cedex 9, France;}
\vs\no

\abstract{On-board the SVOM mission, the Microchannel X-ray Telescope observes the soft X-ray band of the gamma-ray bursts afterglows. The so-called lobster-eye optics focuses X-rays to the camera subsystem that performs imaging and spectroscopy of a region of the sky 58$\times$58 arcmin$^{2}$ wide centered on the burst detected by the ECLAIRs instrument. The recorded photos positions are used by the on-board scientific software to rapidly localize the source, whereas spectral information is used on ground to model the properties of the gamma-ray bursts. The first months in orbit were intensively used to tune the parameter settings of the detector and the calibration method to provide high availability of the camera and accurate spectroscopy to the users. The paper presents the design of the camera validated by on-ground testing, the tuning phase in flight and the performance of the camera at the beginning of the mission. Perspectives are given concerning the evolution of the spectral response during the mission.
}

   \authorrunning{M.~Moita et al.}
   \titlerunning{In-flight performance of the MXT camera}  
                                    
   \maketitle
%
%

\section{Introduction}
\label{sect:intro}

This paper presents the camera subsystem of the Microchannel X-ray Telescope (MXT) on board \textit{SVOM}. The MXT camera is responsible for the imaging and spectroscopy of gamma-ray bursts (GRBs) in the soft X-ray band (0.2--10~keV), following the initial source localisation provided by \textit{ECLAIRs} and the subsequent slew that places the target within the MXT field of view. The imaging capability is used to refine the GRB position, while the spectroscopic information enables the astrophysical interpretation of the afterglow, in particular for the modeling of radiation processes and the environment around the central engine of the burst. For a global description of the full instrument, we refer the reader to Götz et al. (this issue).

The details of the camera design are described in Section~\ref{sect:camera}, together with the results of the ground calibration campaign that supported the first months of in-flight operations. Section~\ref{sect:tuning} presents the tuning of the camera parameters performed during the commissioning phase. Finally, Section~\ref{sect:performance} summarises the in-flight performance of the MXT camera.

\section{Presentation of the camera and ground calibration}
\label{sect:camera}
\subsection{Camera Design and Operation}

\begin{figure*}
\centering
\includegraphics[width=0.8\textwidth]{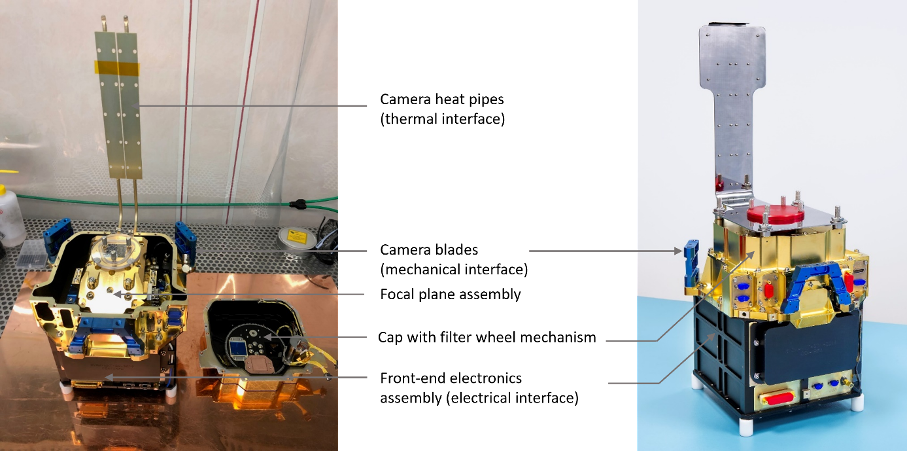}
\caption{Overview of the camera main elements with pictures of the qualification model. (Left) Open configuration to see the filter wheel assembly and the focal plane assembly (b) Fully integrated configuration with a protection on the heat pipes for transportation.}
\label{fig:camera_overview}
\end{figure*}

The camera subsystem consists of an optical bench that hosts the focal plane assembly and supports the thermally decoupled front-end electronics assembly, as shown in Figure~\ref{fig:camera_overview}. The focal plane assembly includes a ceramic board with a $256 \times 256$ pixel pn-CCD read by two CAMEX ASICs, mounted on a system of thermo-electrical coolers and enclosed in a 3~cm thick aluminium shielding. The assembly is connected to propylene heat pipes, which interface with the instrument heat pipes to evacuate the heat load towards the instrument radiator. This combination of passive and active cooling allows the detector to be maintained at $-65^\circ$C with a temperature stability better than 1~degree, even though the camera interface temperature can vary between $-60^\circ$C and $-35^\circ$C (as measured after one year in orbit).

The camera cap is equipped with a wheel that can be rotated into four positions: two open ones, a closed, and source one. The closed position comprises a 1~cm thick piece of copper, which has a stopping power for protons equivalent to that of the permanent shielding outside the field of view. This position is mainly used during regular passages through the South Atlantic Anomaly to minimise displacement damages to the detector caused by trapped protons and their induced secondary particles. The source position incorporates a sealed source of  \({}^{55}\mathrm{Fe}\) with a beryllium window from the Ziegler and Eckart Company. The initial activity was limited to 1~MBq due to export control issues. Using fluorescence targets was considered to produce several lines in the MXT energy range (e.g. Al~K$\alpha$ at 1.5~keV and Ti~K$\alpha$ at 4.1~keV), but the intensities would have been around 500 times lower than the Mn~K$\alpha$ line. Therefore, the direct illumination configuration was preferred. Consequently, the MXT calibration source produces lines at 5.89~keV (Mn~K$\alpha$), 6.49~keV (Mn~K$\beta$), and a much fainter one at 4.74~keV (Si~K$\alpha$ fluorescence escape).

The front-end electronics (FEE) assembly is a stack of three boards to supply the various bias voltages of the CAMEX and the pn-CCD, to encode the analogue outputs of the CAMEX in arbitrary digital unit (ADU), to control the frame readout with a FPGA-based sequencer (Field-Programmable Gate Array–based sequencer), and to perform the first level of data processing and necessary data filtering. The FPGA is also in charge of the communication with the MXT data processing unit (MDPU). The pn-CCD integrates an image area and a frame store area, like the next generation pn-CCD successfully operated in space in the eROSITA instrument on-board the SRG mission (\cite{meidinger06}) and the FXT on-board the Einstein probe mission (\cite{chen2020fxt}. The design of the MXT detector is actually anterior to these missions and was initially produced for the DUO mission project (\cite{meidinger2004duo}). The use of this frame store area allows a significant reduction of out-of-time events (photons impinging the image area during the readout) compared to the former generation of devices in the EPIC camera of the XMM-Newton mission: the image area is quickly transferred (200~µs for MXT) in the frame store area, shielded to X-rays, and is then immediately available for a new photon integration. Following this transfer, the CAMEX ASICs perform the readout of the framestore area, row by row with columns in parallel (one ASIC channel per column, 128 channels per ASIC), which takes 8 ms. The image integration time is fixed to 100~ms: it is well adapted to the maximum photon flux per pixel and it reduces by a factor of 10 the average power consumption of the ASICs (switched on only for 10 \% of the time), as needed for the thermal management of the focal plane. With a pixel signal encoded in a 16-bit word, the frame read-out data rate is 10~Mbits/s. The electrical architecture of the MXT detection chain is designed to perform as many online data-correction steps as possible, thereby simplifying the processing required downstream by the MDPU processor (Robinet et al. (this issue)). These operations had to be elementary and executable within 100 ms, including data transfer. This constraint led to the implementation of two camera readout modes. The nominal mode is the “Event” mode, in which the camera transfers only a (relatively short) list of hit pixels. The “Full Frame” mode, by contrast, transfers all pixels to the MDPU. Its integration time is limited to 200 ms to comply with the data-rate capability of the SpaceWire link between the camera and the MDPU (5 frames per second instead of 10). This mode is used on board for routine parameter settings (baseline level and dark noise, defined below and in Robinet et al., (this issue)) and for on-demand diagnostic activities.

\subsection{Data Processing}

From the full frame data acquired, the first processing steps performed in the FEE consist of applying:

\begin{enumerate}[label=(\roman*),ref=\roman*]

\item A correction of the pixel baseline, $B$: this level (in ADU) corresponds to the average amplitude measured by the system when the pixel is not illuminated. Baseline pixel maps are obtained from dark measurements, i.e. the acquisition of full-frame data when the wheel is in the closed position. Given a succession of dark frames $S$, where a pixel at column $i$ (RAW~Y) and row $j$ (RAW~Z) in frame $k$ is denoted by $S_{ijk}$, the baseline map is defined as the pixel-by-pixel median along the temporal direction:

\begin{equation}
B_{ij} = [S_{ijk}]^{50\%}_{k},
\end{equation}

where $[\cdot]^{50\%}_{k}$ denotes the median computed along the time axis (i.e. over frame number), as frames are acquired sequentially. The correction consists of subtracting this baseline value from each pixel. The dispersion around this median value defines the pixel noise level, referred to as the \textit{dark noise} and is initially expressed in ADU rms. 

\item A row-wise common-mode correction, $CM$  : the median along each row of a baseline-subtracted frame is computed and subtracted from the the pixel amplitudes to compensate for possible common-mode noise during the readout caused by small fluctuations of the bias voltages. 

\begin{equation}
\mathrm{CM}_{ik} = [S_{ijk} - B_{ij}]^{50\%}_{j},
\end{equation}

Each CAMEX can have different common-mode noise, so the correction is actually computed separately for each side. This step has been demonstrated to improve spectral data quality in XMM-Newton and is implemented in MXT for the same purpose.

\item A low-level threshold for each pixel : the \textit{residual amplitude} ($A_r$) is obtained after subtracting the pixel baseline and the row common-mode, is compared to a threshold value defined at 4 times the pixel dark noise level. All pixels whose amplitudes are above their thresholds are defined as hit pixels. This factor 4 is a good compromise between throughput and spectral resolution. Due to the limited number of operations affordable in 100\,ms with the chosen US-free FPGA (ATMEL 480\,FT) running at 20\,MHz, it was not possible (as with other systems based on X-ray detectors) to apply two thresholds or to output neighbor pixels data for this step. This would require accessing the frame content more than once. However, this process was validated during ground calibrations and showed excellent spectral performance \cite{CERAUDO2020164164}.

Other corrections to reduce the amount of data such as high-level threshold or clustering of hit pixels have been considered. However, the gain in data transfer rate have been found minimal compared to the extraction of hit pixels (7 kbits/s for 100 hit pixels corresponding to a bright source). Therefore, the next processing steps have been implemented in the MDPU science software (Robinet et al.\ (this issue)) and in the on-ground pipeline analysis (Maggi et al.\ (this issue)).
\end{enumerate}

\subsection{Results from the Ground Calibration Campaign}

The MXT instrument was calibrated before launch during a three-week campaign in the PANTER X-ray facility in Neuried (Germany)(\cite{schneider23}). During this campaign, the spectral calibration of the MXT camera was performed with X-ray fluorescence lines from 277 eV (C K) to 8910 eV (Cu K$\beta$). The calibration parameters measured in this campaign were used during the first months of operation of the camera in orbit. They consist of a list of 256 column-specific gain and offset values and one detector-global and energy-dependent function of charge transfer inefficiency (CTI) used to compute the energy $E$ (in eV) deposited in a hit pixel from the amplitude $A_r$ (in ADU) measured in this pixel.

The energy before CTI correction is computed using:
\begin{equation}\label{eq:eprime}
  E' = G_i \times A_r + O_i
\end{equation}

To obtain the final corrected energy $E$, the CTI correction is applied as:
\begin{equation}\label{eq:ecti}
  E = \frac{E'}{(1 - \text{CTI(}E'\text{))}^{255 - j}}
\end{equation}

Where:
\begin{itemize}
  \item $A_r$ is the residual amplitude of the hit pixel after baseline and common-mode corrections;
  \item $E'$ is the energy in eV before CTI correction;
  \item $G_i$ is the gain of the column $i$ of the hit pixel in eV/ADU;
  \item $O_i$ is the spectral offset of the column $i$ of the hit pixel in eV;
  \item $255 - j$ is the number of raw transfers to the frame store, with $j$ the row number of the hit pixel;
  \item CTI is the following energy dependent empirical function:
  \begin{equation}\label{eq:cti}
  \text{CTI}(E') = 4.63 \times 10^{-5} \times {E'}^{-0.42}.
\end{equation}
\end{itemize}
This campaign was also very useful to consolidate the method to inter-calibrate events of different multiplicities. X-ray photons can generate, by charge sharing, clusters of 1 to 4 neighbouring hit pixels contained in a matrix of $2 \times 2$ pixels (pixel size of 75~$\mu$m). After summing the energy content of the cluster, the measured photon peak position depends on the multiplicity. The 4-pixel cluster configuration corresponds to the situation where the full energy is measured, whereas the other configurations are likely to be a shared event whose signal is below the threshold for one or several pixels (the average pixel threshold is 45~eV at the beginning of life). Since the contents of the neighbour pixels are not accessible in the output data, the phenomenon can result in incomplete signal collection. This charge sharing effect (CS) and subsequent statistical charge loss were modelled based on experimental data, and CS coefficients are applied to the final energy depending on the multiplicity of events (see \cite{schneider23} for details). 

At the end, with the full data exploitation of the calibration campaign, we measured an on ground energy resolution of 73~eV full width at half maximum (FWHM) at 1.5~keV with single-hit events, and 80~eV FWHM with all event multiplicities; we also found an absolute energy scale accuracy better than 20~eV in the 0.2 to 6.5~keV energy range (\cite{schneider23}).

\section{Tuning of the Camera Parameters}
\label{sect:tuning}

\subsection{Effects of the Environment on the Detector Behaviour}
\subsubsection{Sensitivity to single events effects (SEE)}

\begin{figure}[h]
\centering
\includegraphics[width=0.4\textwidth]{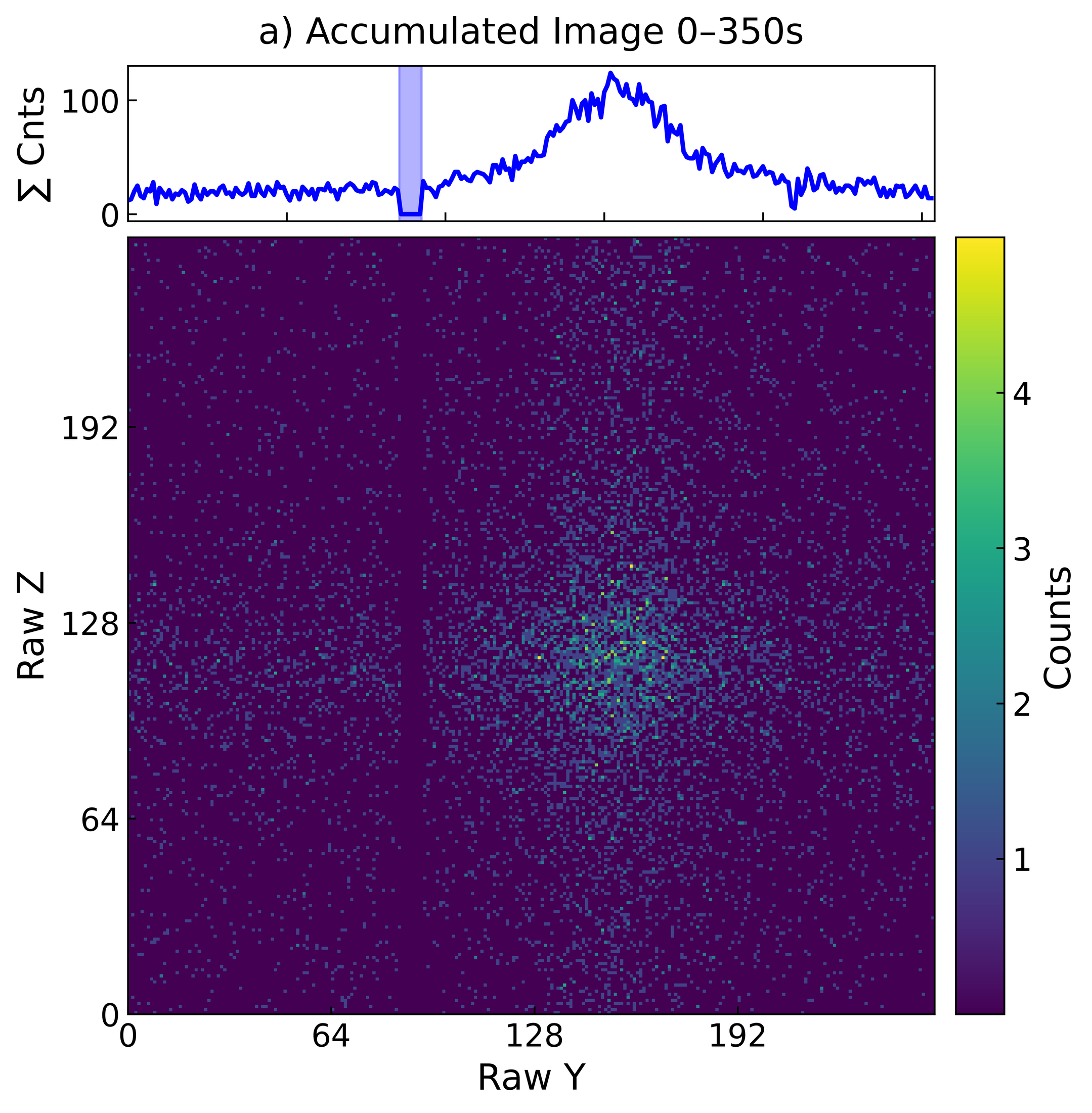}
\hfill
\includegraphics[width=0.4\textwidth]{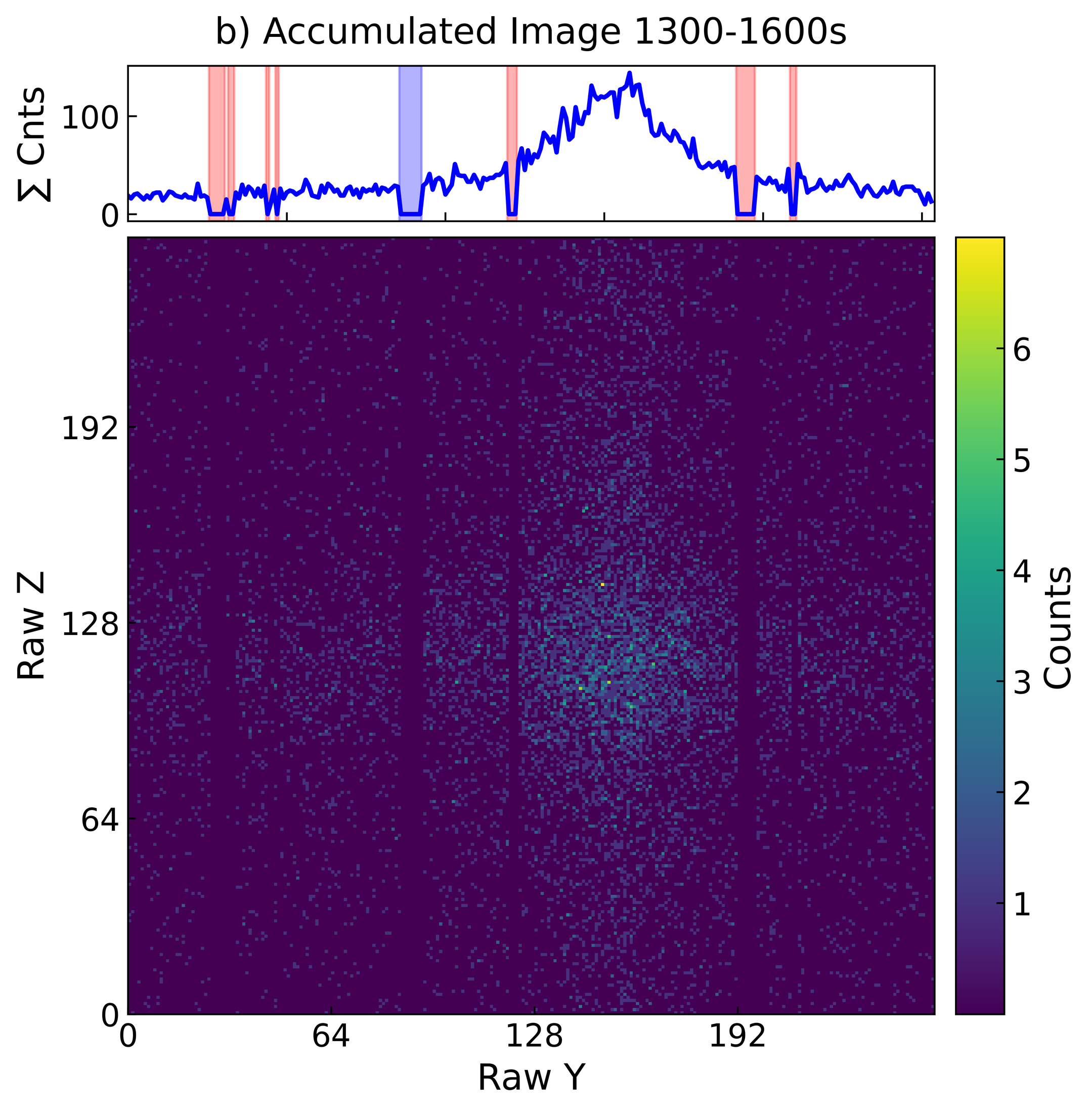}
\caption{Count maps from the same source observation obtained during the first days of MXT operations. Each panel includes, on top, the corresponding column-profile plot. The top panel shows events integrated over 0--350~s after detector switch-on, while the bottom panel corresponds to 1300--1600~s. Several columns become blind at later times, appearing as distinct vertical bands of reduced signal in the images and as sharp drops highlighted in the column profiles.}
\label{fig:columns_loss}
\end{figure}

The detector, initially operated in flight with the same bias settings used during the on-ground calibration campaign, soon exhibited significant temporal instabilities. Figure~\ref{fig:columns_loss} illustrates this behaviour: the top panel shows an event map integrated over 0--350~s after detector switch-on, whereas the bottom panel corresponds to 1300--1600~s. Over this period, several columns clearly became insensitive to photons. This effect is also visible in the column-profile plots displayed above each image, where the insensitive columns appear highlighted. Analysis of the earliest observations showed that these column losses are correlated with the interaction of cosmic rays with the pn-CCD.

This behaviour originates in the internal architecture of the DUO pn-CCD. Each detector column is connected to a first amplification stage within the sensor, implemented as an integrated junction field-effect transistor (hereafter the first FET). A large charge deposition can drive the first FET into saturation and an improper biasing state, preventing the expected charge-to-voltage conversion. As a result, all pixels in the affected column fail to be amplified and the column becomes effectively blind to incoming photons.

The full recovery of the detector active area is possible by switching off and on the detector, at the cost of a period of unavailability for the instrument. The phenomenon (non destructive single event effect) was qualitatively known and its monitoring was anticipated: the sum of the currents drawn by all first FETs is sent in the housekeeping data as well as broadcasted in the recurrent VHF messages. Non-functional first FETs produce a measurable current drop proportional to the number of lost columns, which allows quantifying the insensitive columns even without an external X-ray source. Since the focal plane was also programmed to be switched off during South Atlantic Anomaly (SAA) passages, these insensitive columns could be restored at each SAA passage.

However, the sensitivity of the first FETs to cosmic rays in SVOM’s orbit, outside the radiation belts, had not been characterised on ground and was found to be unexpectedly high: up to half of the detector could be lost within a few tens of minutes. Applying a highly positive voltage to the reset anodes surrounding the first FET mitigates this blocking effect, at the expense of spectral performance though (Sect.~2.2). The commissioning phase was used to optimise this voltage. Given the critical role of MXT in the follow-up of gamma-ray bursts, such column losses, reducing the effective area and introducing potential biases in source localisation, were unacceptable. A sufficiently high reset-anode voltage was therefore adopted to ensure stable operation.

\subsubsection{Sensitivity to Earth albedo flux}

\begin{figure}[h]
\centering
\includegraphics[width=0.4\textwidth]{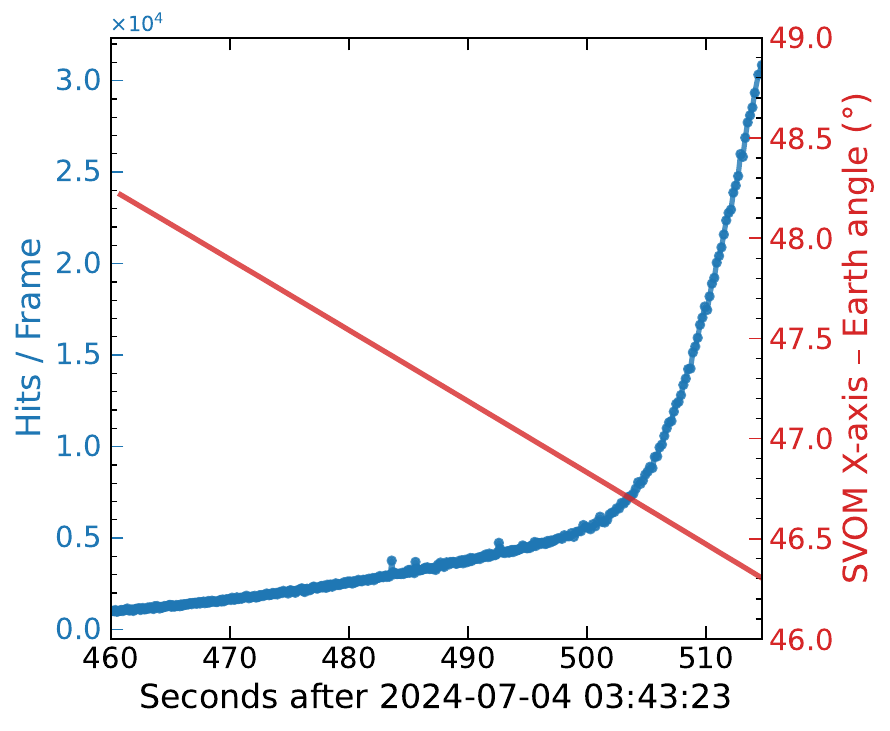}
\hfill
\includegraphics[width=0.4\textwidth]{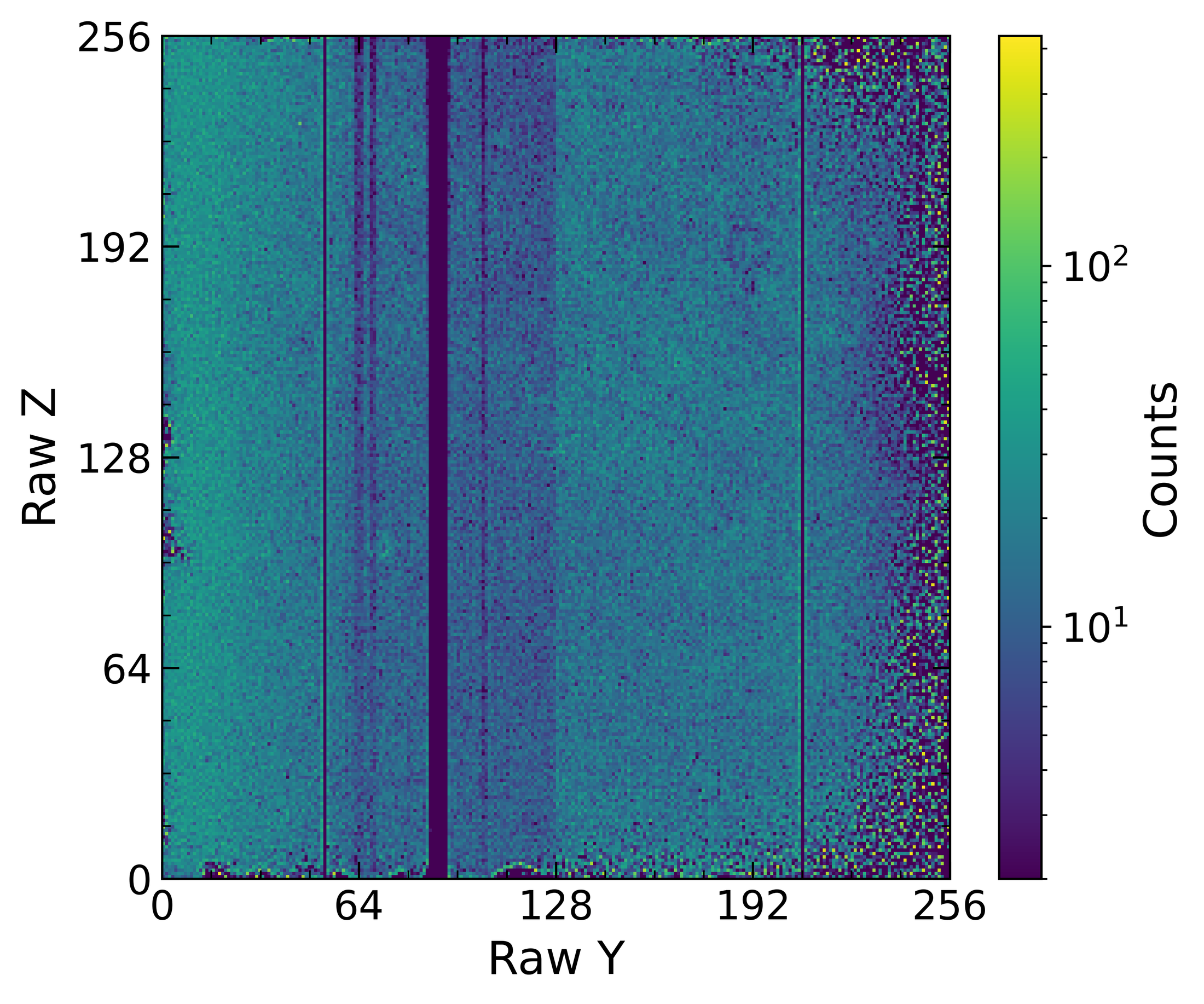}
\caption{(top) Counts per frame and angle between SVOM X-axis pointing and Earth during a straylight event. (bottom) Cumulated image during this passage. Some areas are saturated with events, some columns are empty of events revealing blocked first FET.}
\label{fig:straylight}
\end{figure}

The in-flight camera validation was also significantly constrained by \textit{straylight} contamination. When the Earth approached the instrument’s line of sight, the detector became saturated due to the detection of visible and infrared photons. Pre-launch analyses had defined a lower limit of $20^{\circ}$ on the angle between the SVOM X-axis pointing direction and the Earth limb to prevent such events.  However, early observations showed that straylight contamination can occur at substantially larger angles. As illustrated in Fig.~\ref{fig:straylight}, a pronounced straylight event was recorded even with an X-axis--Earth angle of approximately $47^{\circ}$, well outside the expected exclusion region.

The camera wheel does not include an optical blocking filter, as the sensor itself is coated with a 100~nm thick aluminum layer. The fabrication process was later improved for the detectors on board eROSITA, and Einstein Probe/FXT to achieve the expected transmission of $\sim 10^{-6}$. However, the sensitivity of the DUO pn-CCD to visible light was discovered late in the project, during instrument calibration, when an average transmission of $(4 \pm 2) \times 10^{-3}$ at 637~nm was measured. This behaviour was attributed to the presence of micro-holes in the on-chip filter.

In flight, the optical transmission of the instrument was found to be even higher than during ground testing (see Götz et al.\ (this issue), for details and interpretation). The detection of straylight photons has a two-fold impact on the system, as illustrated in Fig.~\ref{fig:straylight}: it generates a large number of events that must be processed by the MDPU and, similarly to cosmic-ray interactions, can induce large charge deposits that block the FET and lead to the loss of detector columns. Software mitigation alone is therefore insufficient, and the camera wheel must be kept in the closed position whenever straylight is expected. This configuration must be applied during critical orbital phases, but its duration should be minimized to avoid unnecessary loss of observation time. The optimization of these conditions and their on-board management are detailed in Robinet et al.\ (this issue).

\subsection{Consequences of the New Bias Voltages on the Detector Response}
\subsubsection{Additional row effect}

\begin{figure}[h]
\centering
\includegraphics[width=0.45\textwidth]{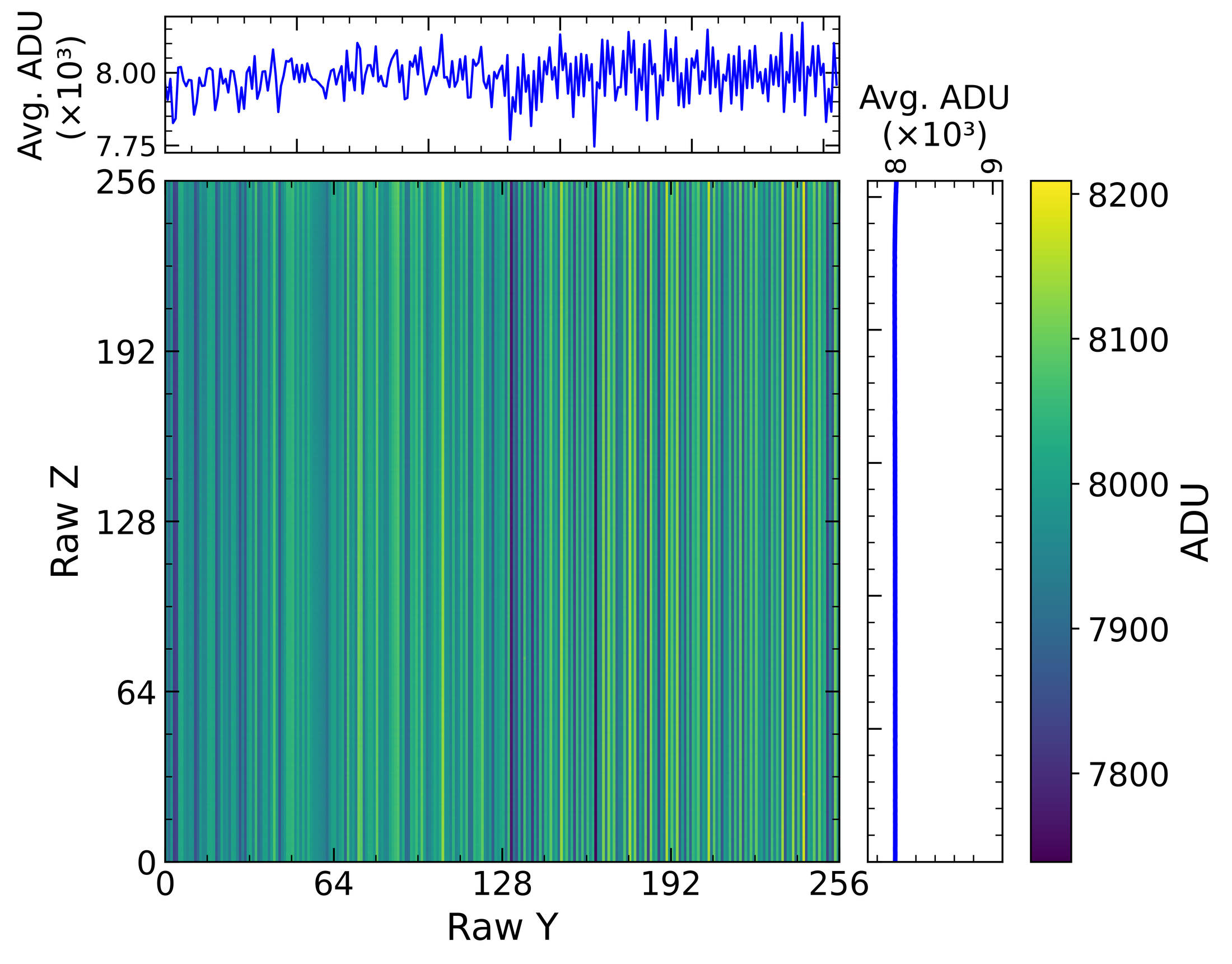}
\hfill
\includegraphics[width=0.45\textwidth]{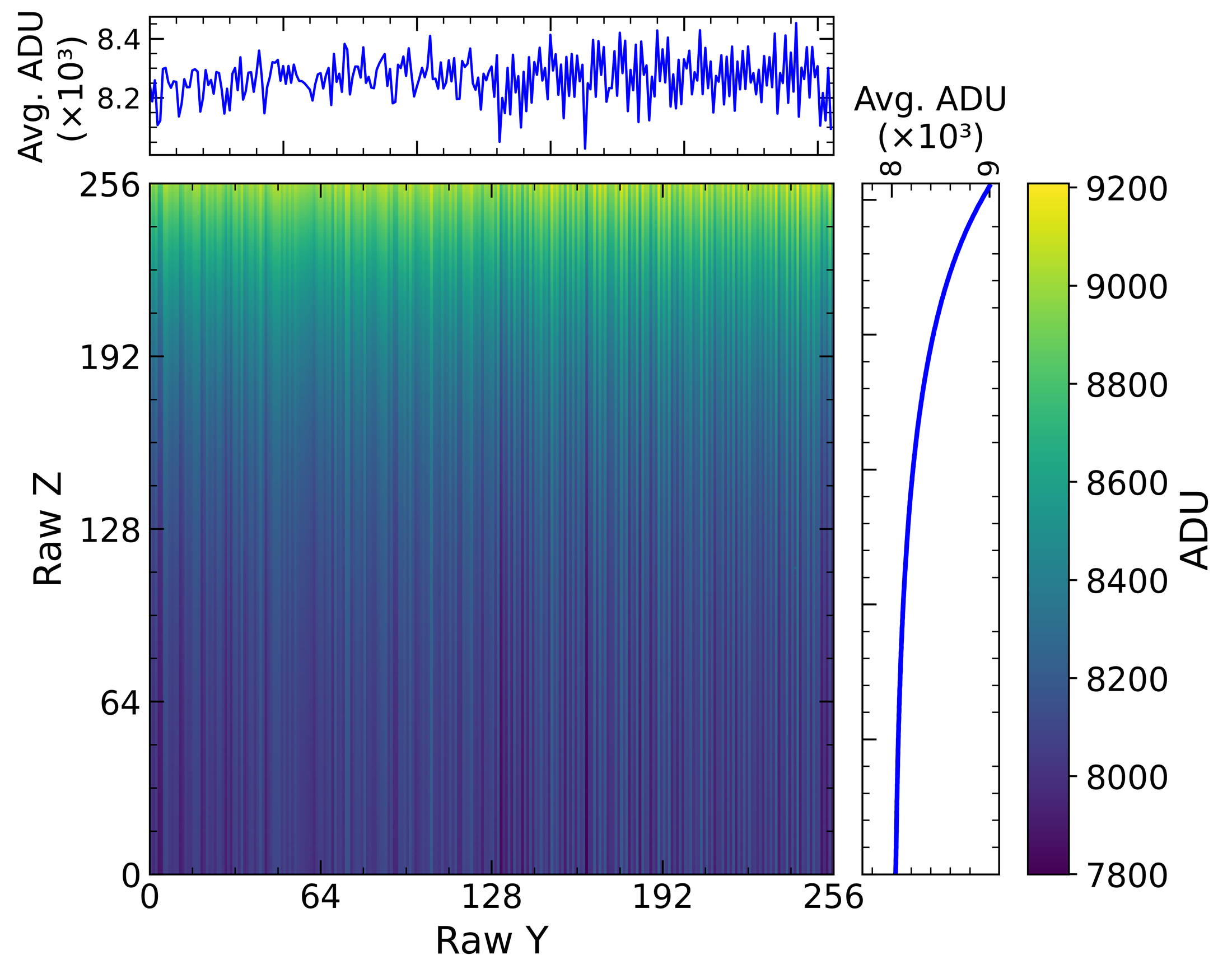}
\caption{Baseline map non-uniformity induced by the change of RSTA voltage: (top) Ground and initial flight configuration (bottom) Final flight configuration.}
\label{fig:baseline_vrsta}
\end{figure}

The updated reset anode (RSTA) voltage introduces a systematic row-dependent pattern, clearly visible in the on-board baseline maps (Figure~\ref{fig:baseline_vrsta}). There is an overall increase of the pixel baseline level, with the effect being strongest in the rows closest to the readout nodes (RAW Z~=~256). 

\begin{figure}
\centering
\includegraphics[width=0.45\textwidth]{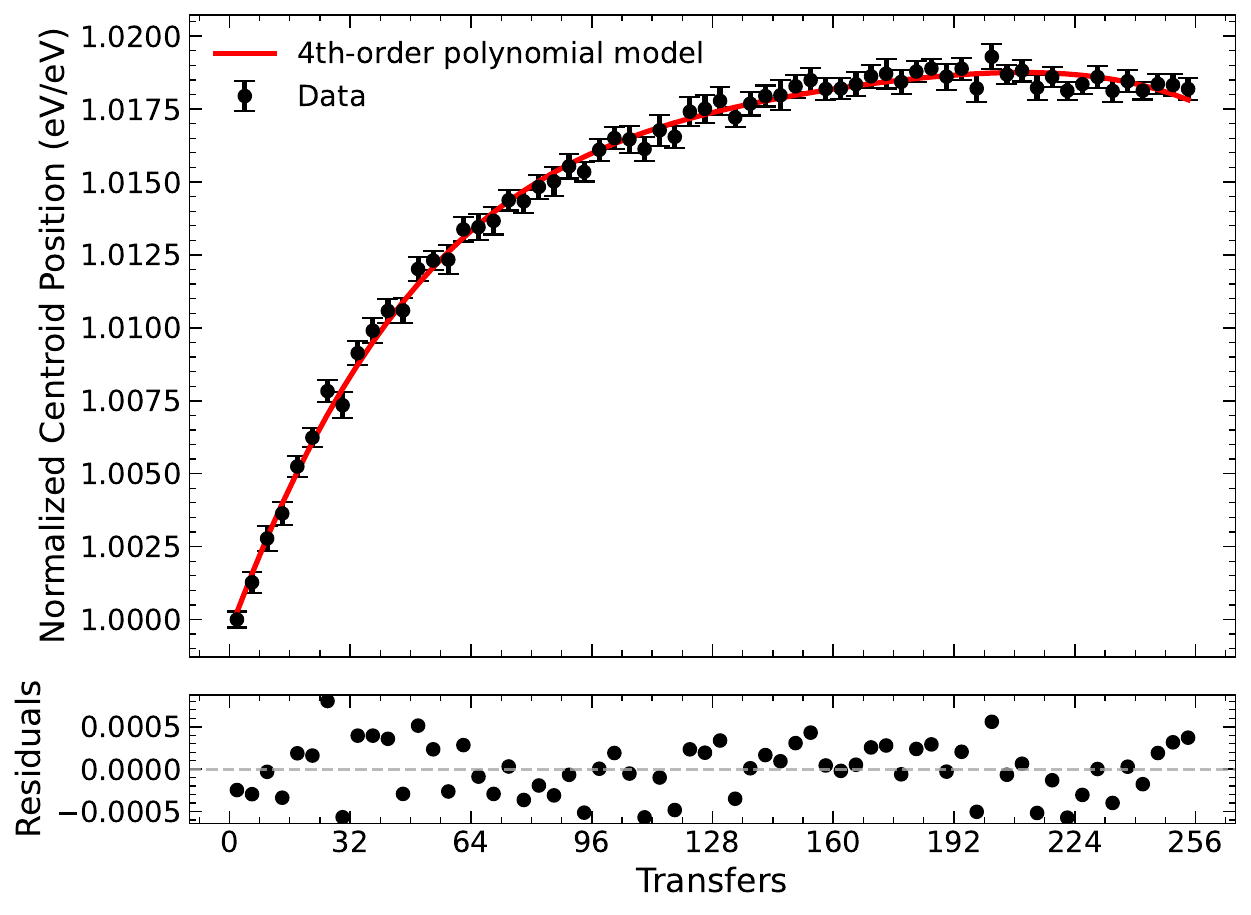}
\caption{Centroid of the 5.9 keV line as a function of the row of the photon interaction, after CTI correction, and modeled by a 4th order polynomial function.}
\label{fig:cti_model}
\end{figure}

Beyond the baseline shift, the event data also reveal a second effect: a pronounced row-dependent modulation of the signal amplitude that persists even after applying both the baseline map and the CTI correction derived from ground calibration. Since the RSTA voltage is pulsed during the readout sequence, it most likely perturbs the analog signals at the first FET stage and within the CAMEX channels. The resulting modulation is highly reproducible and synchronous with the readout, allowing an empirical parametrization with a fourth-order polynomial that is independent of photon energy (Figure~\ref{fig:cti_model}). This correction has been integrated into the MXT ground processing pipeline (Maggi et al., (this issue)), with coefficients regularly validated in flight using the on-board $^{55}$Fe calibration source. The correction remained stable throughout the first year of operations. After one year, measurable deviations from the initial parametrisation were identified and the model was subsequently updated, ensuring that the calibration accuracy is fully maintained.

\subsubsection{Time Variable Low-Energy Electronic Noise}

\begin{figure}[h]
\centering
\includegraphics[width=0.42\textwidth]{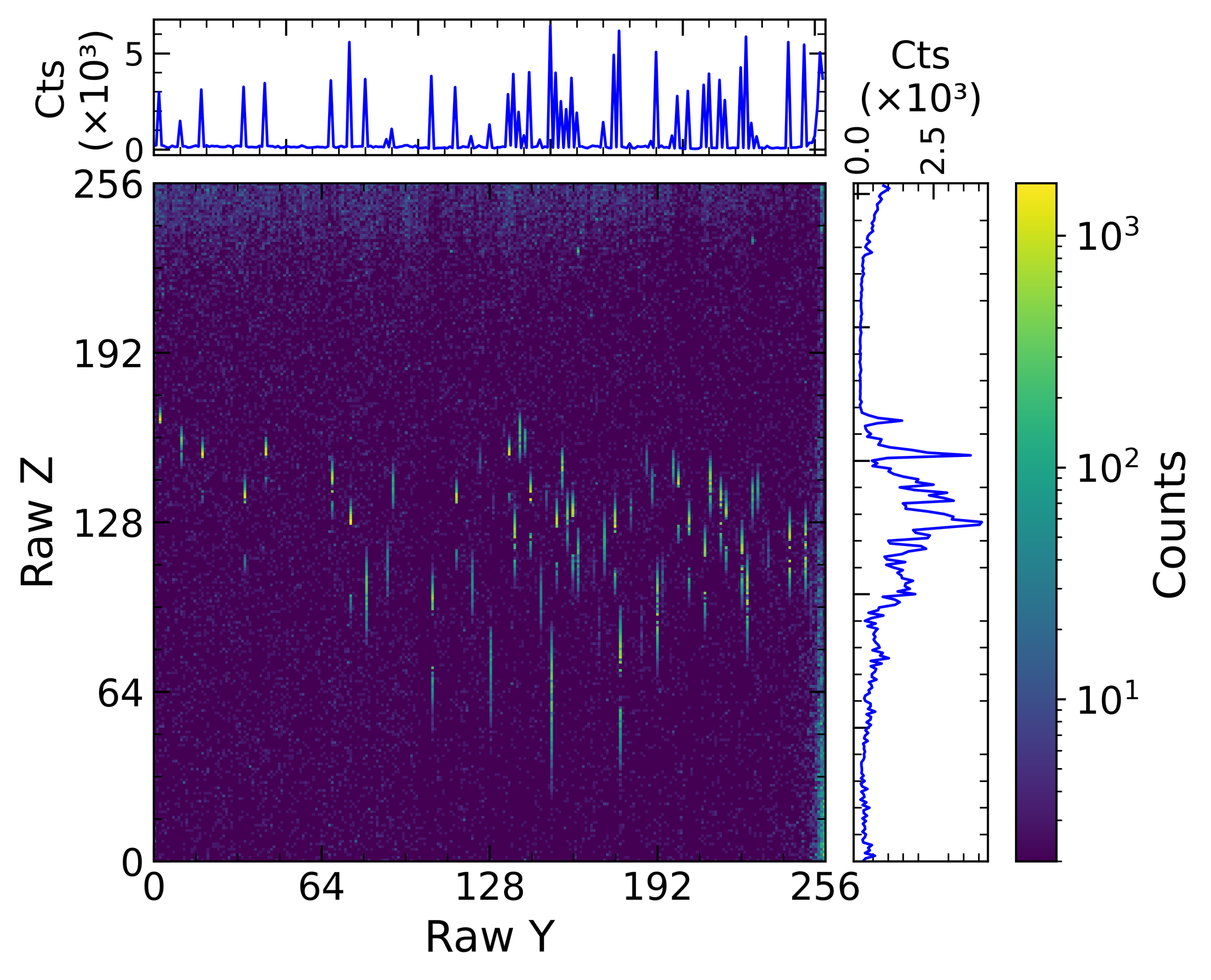}
\hfill
\includegraphics[width=0.4\textwidth]{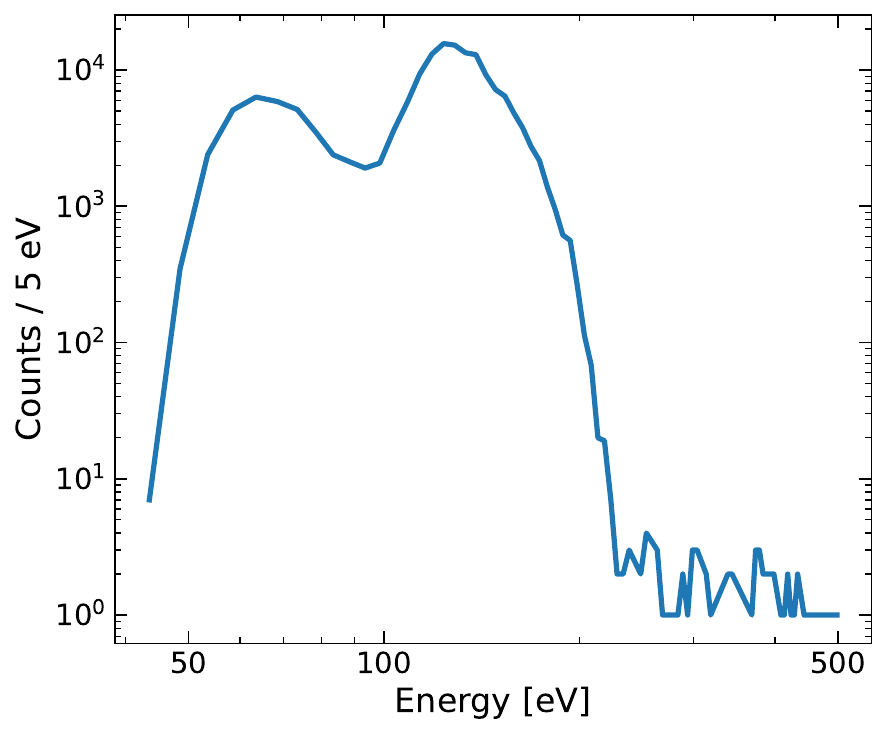}
\caption{Cumulated count map (top) and spectrum (bottom) for energies below 500~eV with the reset anode voltage (RSTA) in the final flight configuration. Both panels correspond to 1000~s of observation with the wheel in source position.}
\label{fig:rain}
\end{figure}

Spurious low-energy events appear as transient bright pixels, producing a characteristic ``rain effect'' when visualizing the cumulative count maps (Figure~\ref{fig:rain}). This effect was not observed during ground tests even with high voltages on RSTA. To mitigate it the on-board localization algorithm applies a uniform 500~eV threshold to all pixels, which efficiently suppresses this effect and prevents any impact on the source position estimate, since less than 10\% of the counts of a typical GRB afterglow are detected between 200 and 500 eV. On ground, a finer treatment is adopted to avoid losing all spectral information at low energies: the bright pixels are identified and treated individually, allowing the spectral information from the remaining pixels to be preserved. Although these events vary with time, they can be reliably identified on an observation-by-observation basis (Maggi et al.\ (this issue)).

\subsubsection{Low-Level threshold computation}

Initially, the computation of Low Level Threshold (LLT) maps was performed onboard using 210 consecutive full frames, corresponding to a total acquisition time of 4.2~s. However, due to the high variability of spurious events  of the "rain effect", this limited dataset, collected over a short period, proved insufficient to accurately identify and exclude noisy pixels. This led to an increase in spurious detections and elevated electronic background, and in worst cases, hot pixels that affected the localisation algorithm, resulting in false detections. To address this issue, LLT maps are now computed on ground using 500 full frames acquired over the course of one week. This extended dataset significantly improves the identification of unstable pixels, allowing for more appropriate threshold adjustments. The updated LLT maps are then uploaded to the instrument once per month. This change has led to a substantial reduction of the electronic background noise.

\subsection{Available Measurements and Calibration Method}

\begin{figure}
\centering
\includegraphics[width=0.45\textwidth]{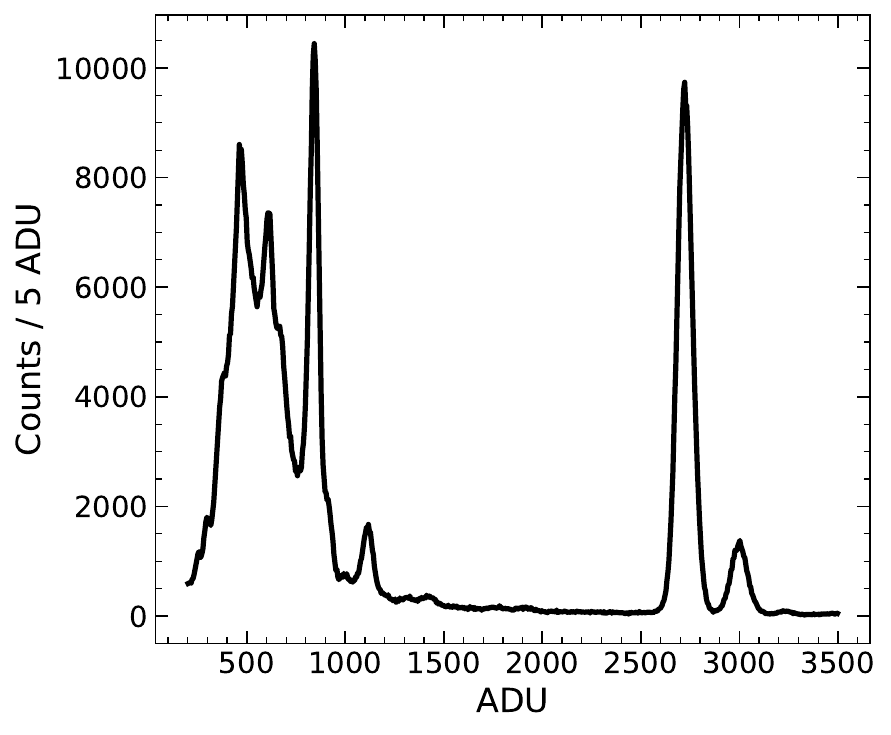}
\caption{Combined spectrum in ADU used for energy calibration, obtained by merging single- and double-pixel events from the same column. The data include Cas~A observations collected between 2024-07-25 and 2024-09-28, together with the on-board $^{55}$Fe source observation from 2024-08-13.}
\label{fig:SpecCalib}
\end{figure}

The calibration of the MXT camera over its full energy range is performed through regular in-flight measurements. The onboard $^{55}\mathrm{Fe}$ source, used every two weeks for 1000~s, provides high-energy calibration data, while the Cassiopeia~A (Cas~A) Supernova Remnant (SNR), observed approximately every six months, characterises the low-energy response. Uniform coverage of all detector is required to obtain sufficient statistics for deriving the calibration parameters. While the onboard source achieves this in a single exposure, Cas~A, because of the MXT PSF shape, requires a strategy of nine pointings, each offset by 15~arcmin to cover the full detector plane, following the procedure used in the PANTER ground calibration campaign \cite{gotz23}. Each Cas~A pointing requires an exposure of about 10~ks to obtain $\sim 500$~counts per column and per row in all relevant spectral lines. All Cas~A observations are combined, together with the closest-in-time $^{55}\mathrm{Fe}$ measurement. The resulting spectrum in ADU is shown in Figure~\ref{fig:SpecCalib}. Producing the calibration parameters involves two main steps: determining the gain/offset values (evaluated column by column) and determining the CTI (evaluated row by row). 

The gain/offset calibration is performed using the Energy Calibration via Correlation (ECC) method \cite{Maier2016}, in its optimised form with adaptive mesh refinement (AMR) to discretise the parameter space \cite{Maier2020}. This technique determines the parameters that maximise the correlation between a synthetic reference spectrum and the un-calibrated observed spectrum (using single events and doubles along the same column). 
The synthetic spectrum is constructed in two steps. First, a spectral model of Cas~A is defined (bremsstrahlung continuum plus Gaussian emission lines for O, Fe, Ne, Mg, Si, S, Ar, Ca, together with a power-law component) based on Chandra/ACIS fits. In parallel, a model of the onboard $^{55}$Fe calibration source is generated.
Both models are then folded through the MXT response, and the resulting spectra are combined with relative normalisations fixed according to previous measurements. The sum of these two response-folded components constitutes the final synthetic spectrum used by the ECC.

The CTI is derived by fitting the centroids of spectral lines using single and double events along the same row \cite{CERAUDO2020164164}. To improve statistics, the fit is performed jointly on four consecutive detector rows. Since CTI is energy dependent, this process is repeated for several energies across the MXT range to produce a CTI(E) relation, which is then applied to the data using Eq.~\ref{eq:cti}.  

Gain and offset are determined column-wise, while CTI is determined row-wise; this makes the two corrections interdependent, and the calibration must therefore be performed iteratively \cite{schneider23}. The first iteration assumes CTI = 0 to obtain an initial set of gain/offset parameters. These are then used to derive an initial CTI estimate, which is applied in the next iteration before recalculating the gain/offset. The process is repeated until the residual CTI, defined as the CTI measured after all corrections, converges to values below $10^{-5}$. Once convergence is reached, the calibration yields the final gain/offset parameters together with the final CTI estimate.

\subsection{Initial In Flight Calibration Parameters}

\begin{figure}[ht]
\centering
\includegraphics[width=0.40\textwidth]{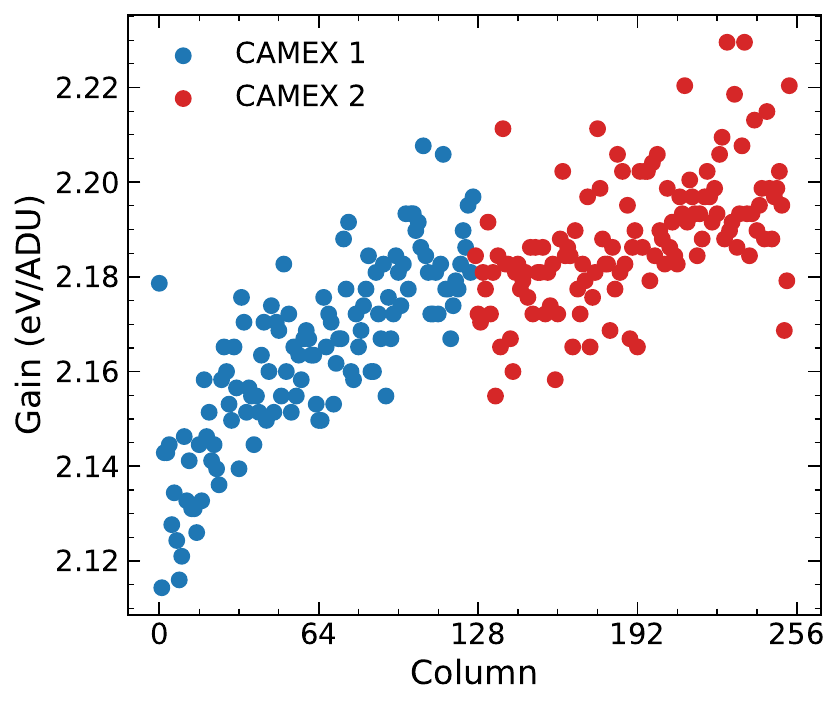}
\hfill
\includegraphics[width=0.40\textwidth]{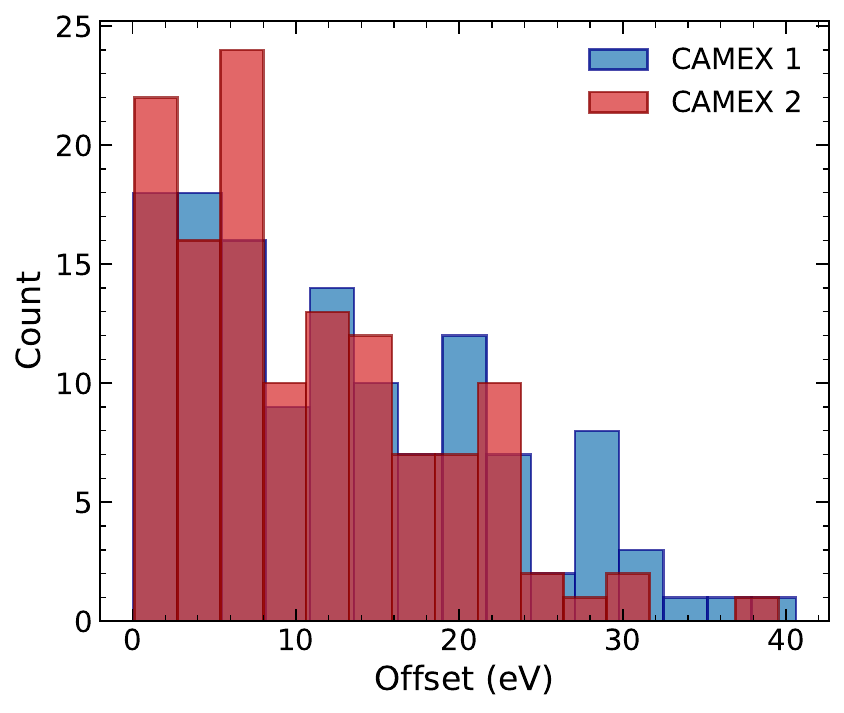}
\caption{(top) Gain as a function of column number.
(bottom) Histogram of the measured offset values.}
\label{fig:gain_offset}
\end{figure}

The gain and offset distributions are presented in Figure~\ref{fig:gain_offset}. We obtain a median gain of 2.2~eV/ADU with a dispersion below 1~\%. The offset shows a median value of 10~eV, but with a relatively large dispersion of about 75~\%.  

\begin{figure}[ht]
\centering
\includegraphics[width=0.42\textwidth]{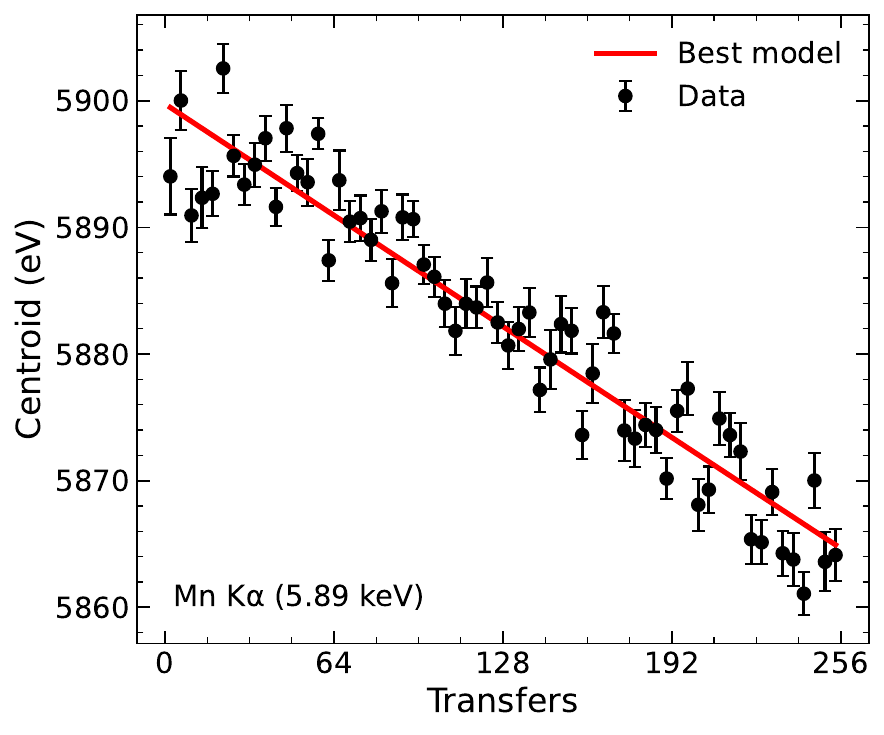}
\hfill
\includegraphics[width=0.40\textwidth]{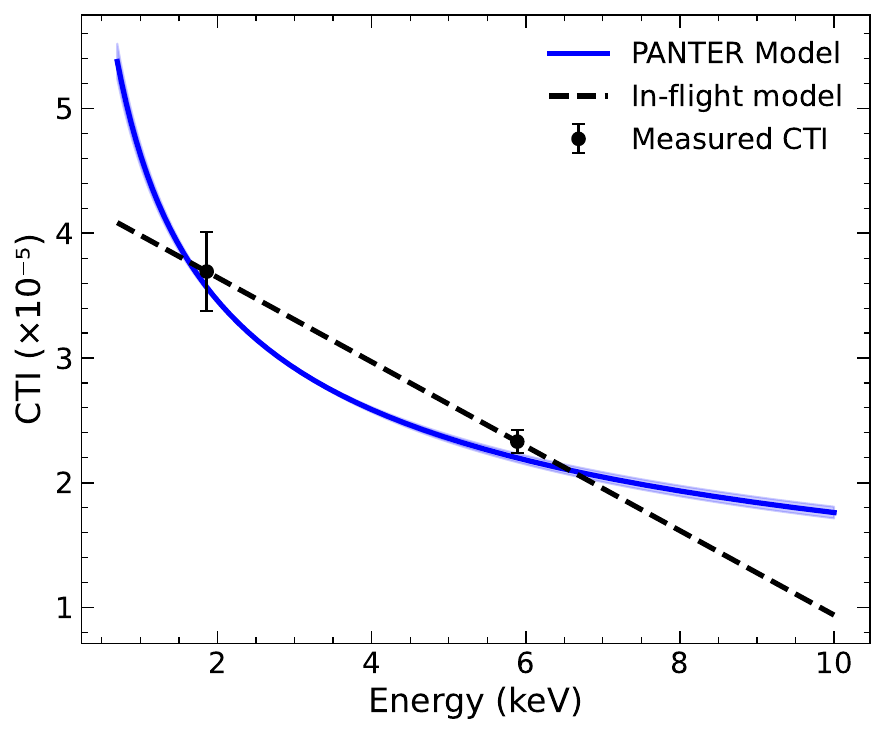}
\caption{(top) Line centroid position of Mn~K$\alpha$ as a function of the number of transfers. Circles indicate the peak positions obtained from Gaussian fits to the row spectra, while the best-fit model used to derive the CTI is shown in red. (bottom) Comparison of the CTI as a function of energy, measured in-flight and on-ground.}
\label{fig:CTI}
\end{figure}

Figure~\ref{fig:CTI} displays the estimated CTI as a function of energy. In the present dataset, sufficient statistics are available for only two spectral lines (1.8~keV and 5.9~keV), limiting the CTI derivation to these energies. Consequently, we performed a linear fit between these two points. For comparison, the figure also shows the CTI model obtained during the PANTER campaign, where the availability of low-energy lines ($<$1~keV) allowed a power-law fit to be used across the full energy range. In the current case, lacking such low-energy lines, we chose not to extrapolate the power-law model so as to avoid overestimating the CTI at lower energies. A linear approximation has been shown to adequately describe the CTI behaviour in previous work \cite{PLASSE2025170717}, and we adopt the same approach here. The best-fit model obtained is:  

\begin{equation}
\label{eq:cti_2}
\mathrm{CTI}(E) =
\bigl[(-3.3 \pm 0.8)\,E + (42.9 \pm 4.6)\bigr]\times 10^{-6},
\end{equation}

with $E$ in keV.

\section{Camera Performance}
\label{sect:performance}
\subsection{In-flight Spectral Performance}

\begin{figure*}[h]
\centering
\includegraphics[width=0.53\textwidth]{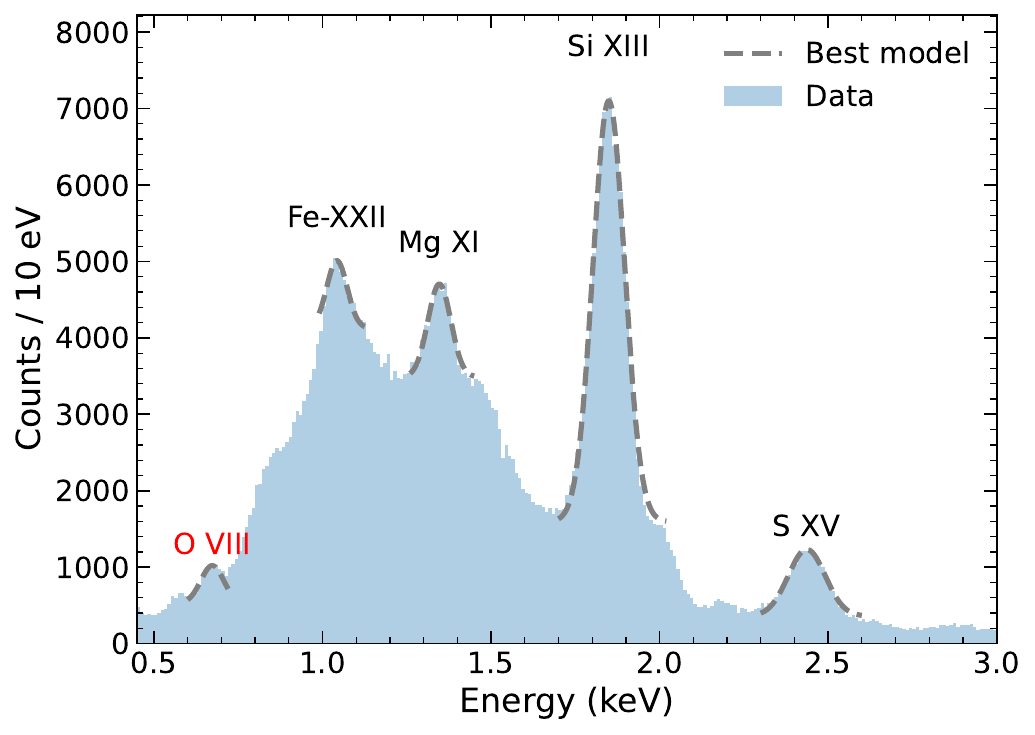}
\hfill
\includegraphics[width=0.38\textwidth]{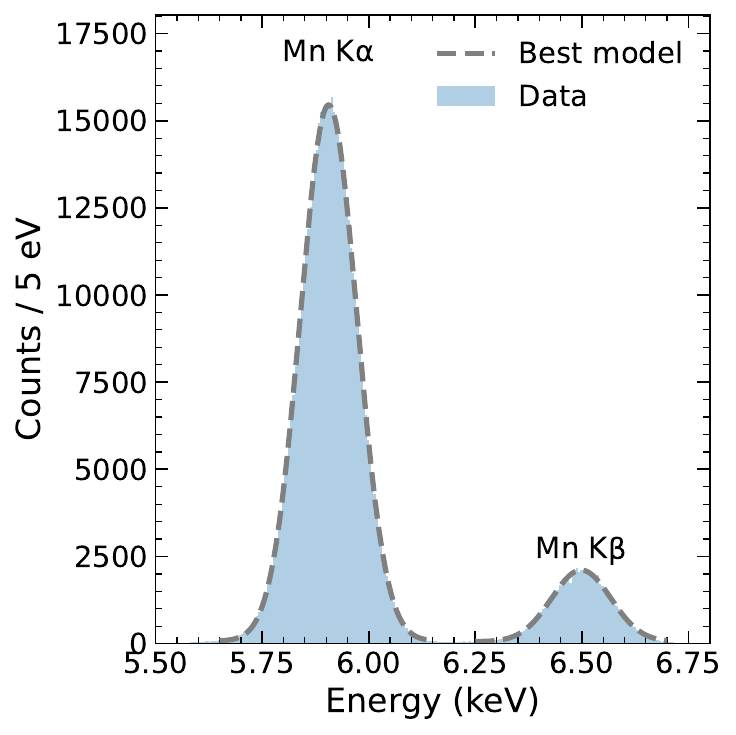}
\caption{Cas A (Left) and $^{55}\mathrm{Fe}$ (Right) spectrum based on August 2024 data and flight calibration. The dashed lines represent the best-fit models used to derive the camera spectral performance.}
\label{fig:spectrum_fits}
\end{figure*}

\begin{figure}[h]
\centering
\includegraphics[width=0.40\textwidth]{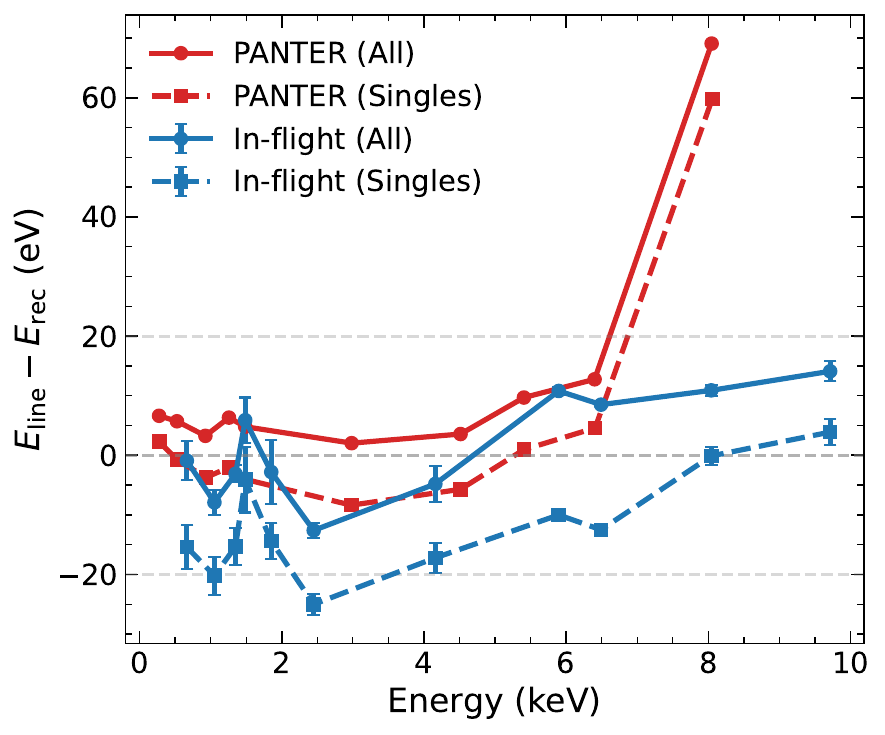}
\hfill
\includegraphics[width=0.40\textwidth]{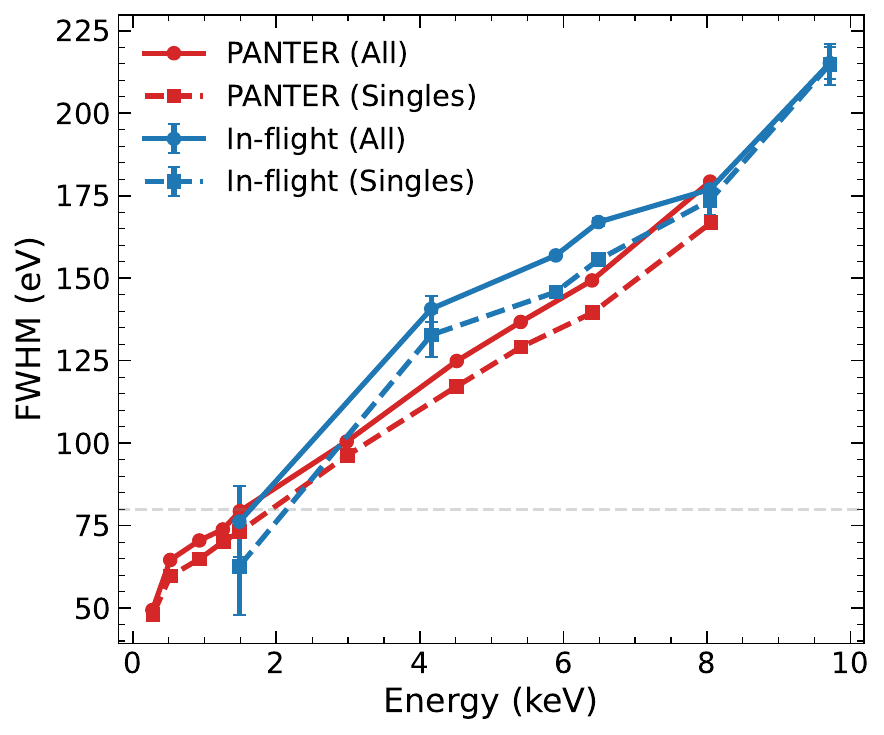}
\caption{Summary of MXT in-flight spectral performance for all event multiplicities and single-hit events (top) Energy plate scale: error on absolute energy. (bottom) Energy resolution at FWHM.}
\label{fig:performance}
\end{figure}

The spectral performance of MXT was evaluated using the on-board calibration source \(^{55}\)Fe (Figure~\ref{fig:spectrum_fits}) and background fluorescence lines (Figure~\ref{fig:bkg}), spanning 1.5--9.7~keV. Each peak was fitted with a Gaussian plus a constant background to extract the centroid position and FWHM. The five most prominent Cas~A emission lines were also used to extend centroid measurements down to 0.6~keV (Figure~\ref{fig:spectrum_fits}). These features, however, are complex blends of multiple transitions and are broadened by plasma dynamics and Doppler velocities, \cite{Plucinsky_2025}. When modelled with a single Gaussian they produce artificially large FWHM values and cannot be used to measure the instrumental resolution. Recovering the true response from Cas~A would require multi-component fitting or plasma models, but this is unnecessary since the instrumental and fluorescence lines already provide reliable constraints on the FWHM for updating the redistribution matrix file (RMF) and ensuring an accurate in-flight spectral response of the detector.

The results are summarized in Table~\ref{tab:mxt_perf}, while Figure~\ref{fig:performance} shows centroid deviations and FWHM as a function of energy, compared with ground measurements at PANTER \citep{schneider23}. MXT meets all spectral performance requirements: using all event multiplicities, the centroid positions remain within \(\pm 20\)~eV across the full energy band, and the Al~K$\alpha$ line at 1.49~keV presents a FWHM of \(76.2 \pm 10.8\)~eV, fully compliant with the 80~eV requirement at 1.5~keV at beginning-of-life.  

These results are in very good agreement with ground measurements and confirm the excellent energy response of MXT in orbit. At the same time, the comparison also reveals a larger discrepancy between line positions for single and all events. This is likely driven by threshold-related charge losses in split-charge events. Although a charge-sharing correction is applied, the residual offset suggests that the adopted model does not fully capture grade-dependent losses under in-flight conditions. In particular, the correction assumes a spatially uniform LLT, whereas in flight, increased time-variable low-energy electronic noise leads to a non-uniform effective threshold. Further refinements of the charge-sharing treatment under the current in-flight conditions are under investigation.

\begin{table*}[h]
\centering
\caption[]{Measured centroid energy and energy resolution for all event multiplicities and single-hit events.}
\label{tab:mxt_perf}
\begin{tabular}{lcl|cc|cc}
\hline
\textbf{Line} & \textbf{Energy (eV)} & \textbf{Origin} &
\multicolumn{2}{c|}{\textbf{Centroid Energy (eV)}} &
\multicolumn{2}{c}{\textbf{FWHM (eV)}} \\
 & & & Singles & All & Singles & All \\
\hline
O VIII       & 665   & Cas A        & 649.6 $\pm$ 3.7   & 664.1 $\pm$ 3.3 & ---  & --- \\
Fe XXII      & 1053  & Cas A        & 1032.8 $\pm$ 3.2  & 1045.1 $\pm$ 2.1 & --- & ---  \\
Mg XI        & 1350  & Cas A        & 1334.7 $\pm$ 3.1  & 1346.9 $\pm$ 1.4 & --- & ---  \\
Al K$\alpha$ & 1486  & Bkg          & 1481.9 $\pm$ 5.5  & 1491.9 $\pm$ 3.8  & 62.7 $\pm$ 14.8  & 76.2 $\pm$ 10.8 \\
Si XIII      & 1852  & Cas A        & 1837.6 $\pm$ 1.0  & 1849.2 $\pm$ 5.4 & --- & --- \\
S XV         & 2449  & Cas A        & 2424.0 $\pm$ 1.8  & 2436.4 $\pm$ 1.2 & --- & --- \\
Si esc       & 4160  & $^{55}$Fe    & 4142.8 $\pm$ 2.5  & 4155.2 $\pm$ 3.0  & 132.8 $\pm$ 6.8  & 140.7 $\pm$ 3.9 \\
Mn K$\alpha$ & 5895  & $^{55}$Fe    & 5885.0 $\pm$ 0.3  & 5905.8 $\pm$ 0.7  & 145.8 $\pm$ 0.7  & 156.9 $\pm$ 0.5 \\
Mn K$\beta$  & 6490  & $^{55}$Fe    & 6477.5 $\pm$ 0.7  & 6498.5 $\pm$ 0.2  & 155.6 $\pm$ 2.0  & 167.0 $\pm$ 1.3 \\
Cu K$\alpha$ & 8041  & Bkg          & 8040.9 $\pm$ 1.5  & 8051.9 $\pm$ 0.9  & 173.6 $\pm$ 4.4  & 176.9 $\pm$ 2.8 \\
Au L$\alpha$ & 9713  & Bkg          & 9716.9 $\pm$ 2.2  & 9727.1 $\pm$ 1.7  & 214.8 $\pm$ 6.3  & 215.2 $\pm$ 4.9 \\
\hline
\end{tabular}
\end{table*}

\subsection{Non X-ray Background}

\begin{figure}
\centering
\includegraphics[width=0.45\textwidth]{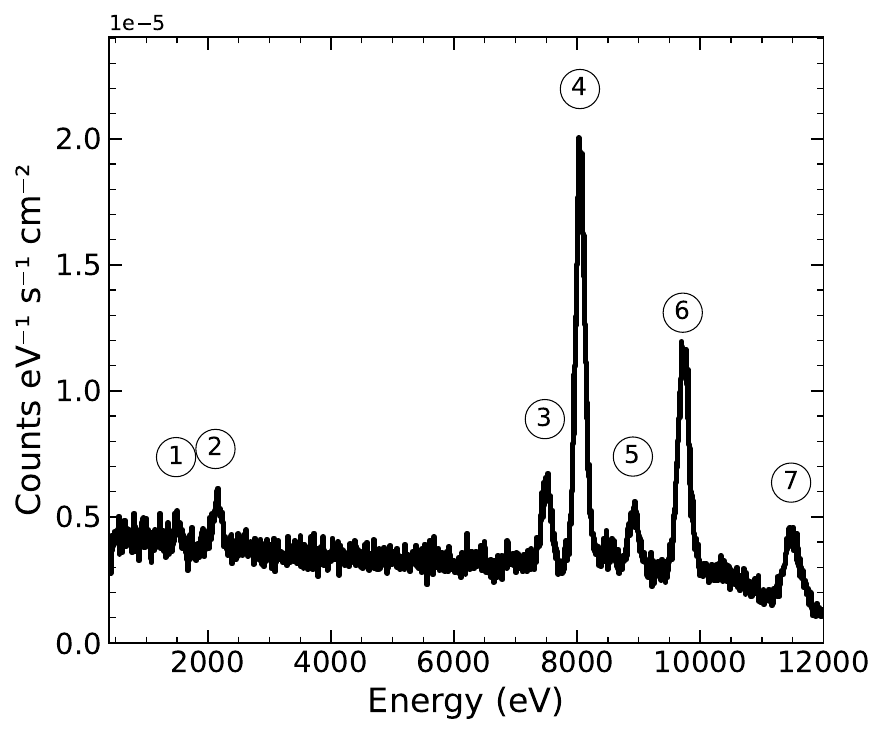}
\caption{Non X-ray background obtained after 30 hour exposure time with the wheel in closed position.}
\label{fig:bkg}
\end{figure}

The instrumental background was measured by switching on the detector with the wheel in closed position and out of the radiation belts. In the energy range of MXT the recorded events mainly correspond to secondary photons generated by the interactions of the cosmic rays with the full satellite and scattered towards the detector. The associated spectrum of cumulated events after energy calibration is shown in Figure~\ref{fig:bkg}. We can clearly identify spectral lines corresponding to X-ray fluorescence of materials in the vicinity of the detector. Except from the entrance cone which is black painted to limit straylight, the inner part of detector shielding is coated with Ni-Au for low thermal emissivity (to reduce the radiative heat flux from the warm environment). This coating was also thought to absorb the X-ray fluorescence from the 30~mm thick aluminum shielding at 1.49~keV. The remaining photons at this energy (line~1) could be due to the 100~nm thick Al on-chip filter covering the full active area of the detector. Copper (line~4) is prominent: it can be found in the MoCu mechanical structure supporting the detector and in the active cooling system made of thermo-electric coolers with interface plates made of copper and pressed on this structure very close to the detector to evacuate the heat load from the CAMEX ASICs. Copper, nickel and gold lines were predicted by Monte Carlo Geant4 simulations in the same proportions and with intensities $\sim$2.5 times lower than those observed. However, the background study was done in the early phases of the project with a preliminary mechanical model of the camera and was not updated during the project. The main motivation of the simulation work was to validate our design strategy to implement background spectral lines that could be used to check for energy calibration but that are not impacting the observations of the X-ray sources in the main energy range of MXT (below 4~keV). It is confirmed by this in-flight background spectrum.

\begin{table*}[h]
\centering
\caption[]{Spectral lines of the camera instrumental background}
\label{tab:bkg}
\centering
\begin{tabular}{cccl}
\hline
\textbf{Line ID} & \textbf{Measured energy (eV)} & \textbf{Line identification} & \textbf{Interpretation} \\
\hline
1 & 1494.2 $\pm$ 7.1   & Al K$\alpha$ (1486 eV)  & On-chip filter \\
2 & 2155.3 $\pm$ 4.3   & Au M$\alpha$ (2122 eV)  & Coating of shielding \\
3 & 7500.9 $\pm$ 3.2   & Ni K$\alpha$ (7472 eV)  & Coating of shielding \\
4 & 8053.0 $\pm$ 1.0   & Cu K$\alpha$ (8041 eV)  & TEC system, MoCu structure \\
5 & 8917.5 $\pm$ 3.8   & Cu K$\beta$ (8905 eV)   & TEC system, MoCu structure \\
6 & 9728.3 $\pm$ 1.7   & Au L$\alpha$ (9713 eV)  & Coating of shielding \\
7 & 11486.9 $\pm$ 4.2  & Au L$\beta$ (11479 eV)  & Coating of shielding \\
\hline
\end{tabular}
\end{table*}

\section{Conclusions}
\label{sect:conclusion}
The first months of MXT operations were dedicated to optimising the detector bias voltages, ensuring both stable performance and a satisfactory spectral response. Within roughly seven months after launch, reference calibration parameters for the energy response were established, combining in-flight measurements from the on-board \textsuperscript{55}Fe source and sky observations of supernova remnants such as Cassiopeia~A to extend the calibration below 4~keV.
As a result of this process, we demonstrated an energy resolution of 80~eV FWHM at 1.5~keV in flight at beginning of life. The current efforts for a continuous progress in the energy calibration precision are focused on our model of charge sharing.
Bi-annual sky-based energy calibration observations are now included in the SVOM plan to maintain the accuracy of the MXT spectral response (energy scale) throughout the mission. Meanwhile, the on-board \textsuperscript{55}Fe source continues to provide bi-monthly monitoring of the 5.9~keV line position (and width), enabling the tracking of charge transfer efficiency degradation and the effects of radiation exposure or solar activity. The long-term evolution of the MXT spectral response will be the subject of a forthcoming publication.

\begin{acknowledgements}
The MXT Camera team warmly thank the MXT project team in CNES (French space agency) for the continuous exchanges during the design and the fabrication of the camera, and the strong involvement during the commissioning phase to rapidly address the anomalies in flight and find viable operational configurations with us (K. Mercier, A. Fort, S. Crepaldi). We also thank MPE (Max-Planck-Institut for extraterrestrial physics) for the procurement of CAMEX ASIC and pn-CDD of excellent performance, for the access to PANTER facility for ground calibrations and for the support to the detector commisionning (N. Meidinger, V. Burwitz). We thank IJCLab and the Observatory of Strasbourg for their flexibility and efficiency to implement new features in the on-board science software and the on-ground analysis pipeline after our observations of some unexpected behaviors in flight of the camera (F. Robinet, P. Maggi).
\end{acknowledgements}

\bibliographystyle{raa}
\bibliography{bibtex}

\label{lastpage}

\end{document}